\documentclass{llncs}
\usepackage{amsmath}
\usepackage{amsfonts}
\usepackage{algorithm}
\usepackage{graphicx}
\def\proof{}
\renewcommand{\proof}[1]{
{\noindent {\it Proof.} {#1} \rule{2mm}{2mm} \vskip \belowdisplayskip}
}

\newcommand{\prevproof}[3]{
{\vskip 0.1in \noindent {\bf Proof of {#1}~\ref{#2}.} {#3} \rule{2mm}{2mm}
\vskip \belowdisplayskip}
}
\newcommand{\vareps}{\varepsilon}

\def\alginp{\noindent{\it Input:~}}
\def\algout{\noindent{\it Output:~}}
\def\algdesc{\noindent{\it Description:~}}

\newlength{\saveparindent}
\newlength{\saveparskip}
\newcommand{\omt}[1]{}

\renewcommand{\Pr}[1]{
{\rm Pr}\left[{#1}\right]
}

\newcommand{\Ex}[1]{
{\rm E} \left[{#1}\right]
}  %

\newcommand{\norm}[1]{||{#1}||}
\newcommand{\ninf}[1]{\norm{{#1}}_\infty}
\newcommand{\none}[1]{\norm{{#1}}_1}
\newcommand{\ntwo}[1]{||{#1}||_2}
\newcommand{\nthree}[1]{||#1||_3}

\def\domain{D}
\def\wmax{w_{\max}}
\def\nsamples{s}
\def\onesixth{\frac{1}{20}}
\newcommand\common[1]{\ensuremath{\mathfrak C(#1)}}
\newcommand\disjoint[1]{\ensuremath{\mathfrak D(#1)}}
\newcommand\sigs[3]{\ensuremath{(#1,#2,#3)}}

\newcommand\numsigs[3]{\ensuremath{|(#1,#2,#3)|}}

\newcommand\numsamples[2]{\frac{\ntwo{#1+#2}}{\ntwo{#1-#2}^2}}

\newcommand{\error}{\frac{polylog{(\nsamples)}}{\nsamples^2}}
\newcommand{\logth}[1]{(\ln #1)^{3/2}}
\title{Telling Two Distributions Apart: a Tight Characterization}
\author{Eyal Even Dar \thanks{evendar.eyal@gmail.com, Final Inc, this research
was performed while author was at Google Research, NY}\and
Mark Sandler\thanks{sandler@google.com, Google Research, NY}}
\institute{}
\def\vareps{\varepsilon}
\def\logo{\tilde{O}}
\begin{document}
\maketitle
\begin{abstract}
    We consider the problem of distinguishing between two arbitrary
    black-box distributions defined over the domain  $[n]$, given
    access to $s$ samples from both.   It is known  that in the worst
    case $\logo(n^{2/3})$ samples is both necessary and sufficient,
    provided that the distributions have  $L_1$ difference of at least
    $\vareps$. However, it is also known that in many cases fewer samples suffice.
We identify a new parameter, that precisely controls the
    number of samples needed, and present an algorithm that requires  the number
of samples only dependent of this parameter and {\em independent} of the
    domain size.  Also for a large subclass of distributions we
    provide a lower bound, that matches our algorithm up to the polylogarithmic bound.
\end{abstract}

\section{Introduction}
\setcounter{page}{0}
\thispagestyle{empty}
One of the most fundamental challenges facing modern data analysis, is to understand and infer hidden properties of the data being observed.
 Property testing  framework~\cite{ron00property,goldreich_propertytesting,batu2000testing,batu2003testing,batu2002complexity,batu2004sublinear} has recently emerged as one  tool to test whether a given data set has certain property with
high probability with only a few queries. One problem that commonly arises in applications is to test whether several sources of random samples follow the same
distribution or are far apart and this is the problem we study in this paper. More specifically, suppose we are given a black box that generates independent
samples of a distribution $P$ over $[ n]$, a black box the generates independent samples of a distribution $Q$ over $[ n]$, and finally a black box that generates independent samples of distribution $T$, which is either $P$ or $Q$. How many samples do we need to decide whether $T$ is identical to $P$ or to $Q$?
This problem arises regularly in change detection problems~\cite{ben2004detecting}, testing whether Markov chain is rapidly mixing~\cite{batu2000testing}, and other contexts.

\paragraph{Our Contribution}
Our results generalize on the results of  Batu et al in~\cite{batu2000testing}, they have shown that
there exists a pair of distributions $P, Q$ on domain $[n]$ with a large statistical
difference $\none{P-Q} \ge 1/2$, such that no algorithm can tell apart case $(P, P)$ from $(P, Q)$ with  $o(n^{2/3})$ samples.
They also provided an algorithm that nearly matches the lower bound for a specific pair of distributions.

In the present paper, instead of analyzing the ``hardest'' pair of distributions, we {\em characterize} the property that controls the number of samples one needs to tell a pair of distributions apart.   This characterization allows us to factor out the dependency on the
domain size.  Namely, for every two distributions, $P$ and $Q$, that satisfy certain technical properties that we describe below, we establish both
an algorithm and a lower bound, such that the number of samples
both necessary and sufficient is $\Omega(\frac{\|P+Q\|_2}{\|P-Q \|_2^2})$. For the lower bound example of~\cite{batu2000testing}, this quantity amounts to $n^{2/3}$  and the distributions in \cite{batu2000testing} satisfy the needed technical properties, and thus our results generalize upon their result.

From practical perspective such characterization is important because the high level properties of the distributions may be learned empirically (for instance it might
be known that the potential class of distributions is a power-law) and our results allow to significantly reduce the number of samples needed for testing.

In many respects, our results complement those of Valiant~\cite{valiant2008}. There it was shown that for testing symmetric and continuous properties, it is both necessary and sufficient to consider only the high frequency elements.  In contrast,  we show that for our problem the {\em low} frequency elements provide all the necessary information.
This was quite surprising provided that low frequency elements have no useful information for continuous properties. For the computability part  ~\cite{valiant2008} introduces the canonical tester - an exponential time algorithm that finds all feasible underlying data, that could have produced the output. If all checked inputs have the property value consistent, it reports it,  and otherwise gives random answer.  In contrast,  our algorithm guarantees that for any underlying pair of distributions, the algorithm after observing the sample will be correct with high probability, even though there might be a valid input that would cause the algorithm to fail.

Finally, we develop a new technique that allows tight concentration bounds analysis of heterogeneous balls and bins problem that might be of independent interest.

\paragraph{Paper Overview}
In the next section we describe our problem in more detail,  connect it to closeness problem studied in the earlier work and state our results.
Section~\ref{sec:preliminaries} proves useful concentration bounds, and introduces the main technique that we use for the algorithm analysis. Section~\ref{sec:algorithm} provides algorithm and analysis, and finally in the section~\ref{sec:lower-bounds} we prove our lower bounds.

\section{Problem Formulation}
We consider arbitrary distributions over the domain $[n]$. We assume that the only way of interacting with a distribution is through a blackbox sampling mechanism.
The main problem we consider is as follows:
\begin{problem}[Distinguishability problem]
Given ``training phase'' of  $s$ samples from $X$ and $s$ samples from $Y$, and a ``testing phase'' of a sample of size $s$ from either $X$ or $Y$,
output whether first or second distribution generated the testing sample.
\end{problem}
We say that an algorithm solves the distinguishability problem for a class of distribution pairs $\cal P$,
with $\nsamples$ samples, if for any $(X, Y) \in \cal P$, with probability at least $1-\error$ it outputs correct
answer. Further, if $X$ and $Y$ are identical, it outputs first or second with probability
at least $0.5-\error$.

We show that the distringuishability problem is equivalent to the following problem studied in \cite{batu2000testing,valiant2008}:
\begin{problem}[Closeness Problem\cite{batu2000testing}]
Given $s$ samples from $X$ and $s$ samples from $Y$, decide whether $X$ and $Y$ are almost identical or far apart
with acceptable error at most $\error$.
\end{problem}
An algorithm solves the closeness problem for a class of distribution pairs $\cal P$, if for every
pair $(X, Y) \in \cal P$, it outputs ``different'' with probability at least $1-\error$, and for every
input of the form $(X, X)$ it outputs ``same'' with probability at least $1-\error$.

Our first observation is that if either of the problems  can be solved for a certain class of distribution pairs, then the other
can be solved with at most  logarithmically more samples and time. The following lemma formalizes
this statement, and due to space limitations the proof is deferred to appendix.
\newcommand{\polylog}{polylog}
\begin{lemma}
\label{lemma:identity-distinguish}
If there is an algorithm that solves {\em distinguishability} problem for a class of distribution pairs $\cal P$ with $s$ then there is an algorithm that solves {\em identity} problem
for class $\cal P$ with at most $O( s \log s)$ samples.

If there is an algorithm that solves {\em closeness} problem
for class {\cal P} with at most $s$ samples, then there is an algorithm that solves
{\em distinguishability} problem with at most $s$ samples.
\end{lemma}

\section{Results}
Our algorithmic results can be described as follows:
\begin{theorem}
\label{thm:alg}
Consider a class of distribution pairs such as $\frac{\ntwo{P-Q}^2}{\ntwo{P+Q}} \ge \alpha$, and let $s = 60(|\log {\alpha}|)^{7/2}/\alpha$,
and each $p_i$ and $q_i$ is at most $\frac{1}{2s}$,  then  Algorithm \ref{alg:main} produces correct answer with probability at least $ 1 - c/s^2$, where $c$ is
some universal constant.
\end{theorem}
Essentially the theorem  abouve states that $\numsamples{P}{Q}$  controls the distinguishability of distribution pair. There are several interesting cases, if $\ntwo{P-Q}$ is comparable to either $\ntwo{P}$ or $\ntwo{Q}$, then the results says that $s \approx 1/ \ntwo{P-Q}$ suffices. This  generalizes the observation from~\cite{batu2000testing} that  if the $L_2$ difference is large then constant number of samples suffices.

We also note that the condition $p_i \le \frac{1}{s}$ is a technical condition that guarantees that any fixed element has expectation of appearance of at most $1/2$. In
other words, no elements can be expected to be seen in the sample with high probability. This is requirement is particularly striking,  given that the
results of~\cite{valiant2008} say that elements that have low expectation, are  provably {\emph not} useful when testing continuous properties.
Further exploiting this contrast is a very interesting open direction.

For the lower bound part, our results apply to a special class of distributions that we call {\emph weakly disjoint} distributions
\begin{definition}
Distributions $P$ and $Q$ are weakly disjoint if every element $x$ satisfies one of the following:
$$(1)~ p_x = q_x, (2)~ p_x > 0 \mbox { and }q_x =0, (3)~q_x >0 \mbox{ and }p_x =0$$
We denote the set of elements such that $p_x=q_x$ by $\common{P,Q}$, and the
rest are denoted by $\disjoint{P,Q}$
\end{definition}

It is worth noting that the known worst case examples of~\cite{batu2000testing} belong to this class. We conjecture that weakly disjoint distributions represent the
\emph{hardest} case for lower bound, and all other distributions need fewer samples, and that the result above generalizes to all distributions.

For the rest of the paper we concentrate on the distinguishability problem, but results through lemma \ref{lemma:identity-distinguish} immediately apply to closeness  problem. Now we formulate our lower bounds results:
\begin{theorem}
\label{thm:lower-bounds}
 Let $P$ and $Q$ are weakly disjoint distributions, and let
$$s \le \min\{ \frac{0.25}{\nthree{P-Q}}, 1/\ninf{P}, 1/\ninf{Q},   \frac{1}{c}\numsamples{P}{Q}\}
$$
where $c$ is some universal constant. No algorithm can solve a distinguishability problem for a class of distritribution pairs defined
by arbitrary permutations of $(P, Q)$, e.g. $\cal P = \{((\pi P, \pi Q)  \}$.
\end{theorem}
The first, second and third constraints in $\min$ expression above are technical assumptions. In fact for many distributions  including the worst-case scenario of~\cite{batu2000testing}
$\numsamples{P}{Q} <<  \min\{1/\nthree{P-Q}, 1/\ninf{P}, 1/\ninf{Q}\}$, and hence those constraits can be dropped.

\section{Preliminaries}
\label{sec:preliminaries}
\subsection{Distinguishing between sequences of Bernoulli random variables}
\label{sec:bernoulli}
Consider two sequences of random Bernoulli variables $x_i$ and $y_i$. In this section we characterize  when the sign of
the observations of $\sum x_i - \sum y_i$ can be expected to be the same as sign of $\Ex{\sum x_i - \sum y_i}$ with high probability. We defer the proof to the appendix.
\begin{lemma}
\label {cor:bernoulli}
 Suppose $\{x_i\}$ and $\{y_i\}$ is  a sequence of Bernoulli random variables such that
$\Ex{\sum x_i} = \alpha$ and $\Ex{\sum y_i} = \beta$, where $\alpha < \beta$.
Then $$\Pr{\sum x_i > \sum y_i } < 2\exp[-\frac{(\alpha-\beta)^2}{8(\alpha+\beta)}]$$
\end{lemma}

\def\domain{[n]}
\def\wmax{w_{\max}}
\def\sample{{\cal S}}
\subsection{Weight Concentration Results For Heterogeneous Balls and Bins }
\label{subsec:weight}
Consider a sample $\sample=\{s_1, s_2, \dots, s_s\}$ of size $\nsamples$ from a fixed distribution $P$ over the domain $[n]$.
Let  the sample configuration be $\{c_1, \dots c_n\}$, where $c_i$ is  the number of times element $i$ was selected.  A standard interpretation is that sample represents
$\nsamples$ balls being dropped into $n$ bins, where $P$ describes the probability of a ball going into each bin. The sample configuration is the final distribution of balls in the bins. Note
that $\sum c_j = \nsamples$.  We show tight
concentration for the quantity $\sum_{i=1}^n  \alpha_i  c_i$, for non-negative $\alpha_i$.\footnote{In fact the technique applies to arbitrary bounded functions.}

Note that $c_i$ and $c_j$ are correlated and therefore the terms in the sum are not independent.
An immediate approach would be to consider a contribution of $i$th sample to the sum. It
is easy to see that  the contribution is bounded by $\max \alpha_i$, and thus one can apply McDiarmid
inequality\cite{mcdiarmid1989method}, however the resulting bound would be too weak since we have to apply an uniform upper bound.

In what follows we call the sampling procedure , where each sample is selected independently from distribtuion over the domain $[n]$ the \emph{type I} sampling. Now,
 we consider a different kind of sampling that we call type II.
\begin{definition}[Type II sampling] For each $i$ in $\domain$ we flip $p_i$-biased coin $\nsamples$ times, and select the corresponding element
 every time we get head.
 \end{definition}
 We show that for almost any sample selection of type (I),  the corresponding configuration in type II sampling would have similar weight in type II.
The weight of all configurations in type ( I) not satisfying this constraint is $o(1/\nsamples)$
$\vareps_1$.
Once we show that, then any concentration bound in Type II will translate to corresponding
concentration bound for type I.

\newcommand{\PRI}[1]{P^{(I)}\left[#1\right]}
\newcommand{\PRII}[1]{P^{(II)}\left[#1\right]}
\newcommand{\config}{{\cal C}}
\def\lprime{{\nsamples'}}
We use $\PRI\cdot$ and $\PRII\cdot$ to denote probability according to type (I) and type (II) sampling.  Due to space limitations all the proofs are deferred to appendix. Now we show the lower and upper bound connections between type I and type II sampling.
\begin{lemma}
\label{lemma:weight-concentration-base}
  For every configuration ${\config} = \{i_1, \dots i_n\}$, such that $i_j \le \ln \nsamples$,
and $\sum i_j = \lprime$, where $\lprime \in [\nsamples \pm \sqrt{\nsamples}]$
   $$
    \frac{2}{3}\sqrt{\nsamples}  \PRII{{\config}}\le \PRI{\config} \le 30 \nsamples^{3/2} \PRII{{\config}}.$$
   where $\PRI{\config}$ is a probability of observing $\config$ in
 type I sampling with $\lprime$ elements and $\PRII{\config}$ is a probability of observing
 \config in type II sampling with $\nsamples$   elements.
\end{lemma}
The following lemma bounds the total probability of a configuration which elements appearing more than
$\ln \nsamples$ times.  The proof is deferred to appendix.
\begin{lemma}
\label{lem:outlier-prob}
In type I sampling, the probability there exists an element that was sampled more than $\ln \nsamples$ times is at most $\frac{1}\nsamples^{\ln \ln \nsamples}$.
\end{lemma}
Now we formalize the translation between concentration bounds for Type (I) and Type (II) samplings.
\begin{lemma}
\label{concentration-lemma}
  Suppose  we sample from distribution $P$,  $s$ times using type $I$ and type $II$ sampling resulting in configurations $C$. Let $A=\{\alpha_i\}$ be an arbitrary vector with non-negative elements, and $r\ge 0$. Then $$
\PRI{|W - \Ex{W}| > r} \le 30 s^{3/2} \PRII{|W - \Ex{W'}| > r} + \frac{1}{s^{\ln\ln\nsamples}},
$$
where $ W = \sum_{i=1}^n \alpha_i c_i$.
\end{lemma}

\begin{lemma}
\label{cor:weight-concentration}
  Consider $s$ samples selected using Type I sampling from the distribution $P=\{p_i\}$,  where  $s \ge 10$,
and $A=\{\alpha_i\}$ is arbitrary vector. Let $W = \sum_{i=1}^n \alpha_i c_i$. Then
$$ \Pr{|W - \Ex{W} | \ge 2 (\ln \nsamples)^{3/2} \ntwo{A}} \le \frac{1}{\nsamples^2}$$
\end{lemma}

\section{Algorithm and Analysis}
\label{sec:algorithm}
At the high level our algorithm works as follows.  First, we check if the two-norms of distributions $P$ and $Q$ are sufficiently far away from each other. If it
is the case, then we can decide  simply by looking at the estimates of the 2-norm of $P$, $Q$ and $T$.
On the other hand if $\ntwo{P} \approx \ntwo{Q}$ then we show that counting the number of collisions of sample from $T$, with $P$ and $Q$, and then choosing the one that has higher number of collisions gives the correct answer with high probability.  Algorithm~\ref{alg:2-norm} provides a full description on $2$-norm estimation. The idea is to estimate a probability mass of a sample $S$ by
computing the number of collisions of fresh samples with $S$, and then noting that the expected mass of a sample of size $l$ is $l\ntwo{P}$.  One important caveat is that
if $S$ contains a particular element more than once, we need to carefully compute the collisions in such a way to keep the probability of a collision at
$l\ntwo{P}$ and to achieve that, we split our sampling into $\max c_i$ phases. During
phase $i$ we only count collisions with elements that have occurred at least $i$ times.
For more details we refer to algorithm~\ref{alg:sampling} which is used as
 subroutine for both~\ref{alg:2-norm} and the main algorithm \ref{alg:main}.
For the analysis we mainly use the  technique developed in Section \ref{sec:preliminaries}.
\paragraph {Sampling according to given pattern}
\def\dOne{P}
\def\dTwo{Q}
\begin{algorithm}[t]
\caption{Sampling according to given pattern}
\label{alg:sampling}
\alginp Configuration $\{c_1, \dots c_n\}$, where   $m \ge c_i \ge 0$. Distribution $P$ \\
\algout Multi-Set of elements $S$, such that $\Ex{|S|} = \sum_{i=1}^n   c_i p_i$ \\
\algdesc
\begin{enumerate}
 \item Sample $m$ elements from $P$, $s_1$, \dots $s_m$
 \item For each $s_i$, if $c_{s_i} \ge i$ include $s_i$ into set $S$
\end{enumerate}
\end{algorithm}
\newcommand{\ntwoapp}[1]{\tilde{\ntwo{#1}}}
\begin{lemma}
 \label{lemma:alg-sampling}
Algorithm~\ref{alg:sampling} outputs a set, $S$, such that $\Ex{|S|} = \sum_{i=1}^n   c_i p_i$.
\end{lemma}
\proof{
 Let $w_i$ be the contribution out of $s_i$.
$
 \Ex{w_i} = \sum_{j=1}^n p_j \mathbf{1_{c_j \ge i}}
$
Summing over all elements of $S$  and use linearity of expectation we have:
$
 \Ex{|S|} = \sum_{i=1}^m  \Ex{w_i} = \sum_{i=1}^m  \sum_{j=1}^n p_j \mathbf{1_{c_j \ge i}} = \sum_{i=1}^{n}p_i c_i,
$
where the last equality follows from changing the summation order.
}
\begin{algorithm}[t]
 \caption{Computing 2-norm of distribution $P$}
 \label{alg:2-norm}
 \alginp Distribution $P$,  accuracy parameter $l$\\
 \algout Approximate value of $\ntwo{P}^2$ \\
 \algdesc
\begin{enumerate}
\item  Select $l$ samples from $P$, let $c_1, \dots c_n$ be the configuration.
 Note that the expected weight $W$ of the configuration is $l \ntwo{P}^2$
 \item If $\max c_i \ge \log l$ report failure
 \item \label{alg:2-norm-sample} Sample with repetition using Algorithm \ref{alg:sampling} $l$ times.
Let $r_i$ be the respective set size from $i$th simulation.
 Note that $\Ex{r_i} = W$
 \item Report $\sum_{i=1}^l r_i / l^2$ as the approximate value
\end{enumerate}
\end{algorithm}
\begin{lemma}
\label{lemma:2-norm-ber}
  The total number of elements selected after $l$ iterations in step \ref{alg:2-norm-sample} is a sum of independent Bernoulli random variables, and its expectation is $l W$.
\end{lemma}
\proof{Indeed, the expected number of selections for every invocation Algorithm \ref{alg:sampling} is $W$,
and that in itself is a sum of $m \leq \log \nsamples$ Bernoulli variables. Thus, we have $lt \log \nsamples$ Bernoulli random variables
with the total expected weight of $lW$.
}
Furthermore, the total number of samples used by the algorithm is bounded.
\begin{lemma}
 The total number of samples used by Algorithm \ref{alg:2-norm} is at most $2 l \log l$.
\end{lemma}
Now we are ready to prove the main property of the algorithm $\ref{alg:2-norm}.$
\begin{lemma}[Concentration results for 2-norm estimation]
\label{lemma:algorithm-2-norm}
 Suppose the Algorithm \ref{alg:2-norm} is run for distributions $P$ and $T$ with parameter $l > 10$.
If $P=T$ then the estimate for $\ntwo{P}$ is greater than estimate for $\ntwo{T}$ with probability $1/2$.
If  $$
\nsamples (\ntwo{T}^2 -\ntwo{P}^2) \ge 4\logth{l} ({\ntwo{P}+\ntwo{T}})
$$
then the estimate
for $\ntwo{P}$ is smaller than the estimate for $\ntwo{T}$ with probability at least $1 - c/l^2$, where $c$ is some universal constant.
\end{lemma}
\proof{
The first part is due to the symmetry. For the second part, we use Lemma~\ref{cor:weight-concentration}  to note
that the weight of selection $ W_T \le s\ntwo{T}^2 - 3\logth{l}) \ntwo{T}$ or $ W_P \ge s \ntwo{P}^2 + 3\logth{l} \ntwo{P}$
with probability at most $\frac{2}{l^2}$. Therefore with probability at least $1- \frac{2}{l^2}$ we have
$ |W_T - W_P| \ge  \logth{l}(\ntwo{P}+\ntwo{T}) $.
Using lemmas \ref{lemma:2-norm-ber} and  \ref{cor:bernoulli} we have:
$$
\Pr{\tilde{W}_T \le \tilde {W}_P} \le o(\frac{1}{l^2}) + 2\exp\left[\frac{l^2(W_T - W_P)^2}{8(W_T + W_P)}\right] \le
\frac{1}{2l^2} + 2\exp \left[ -\frac{l^2(\ln^3 l)(\ntwo{P}+\ntwo{Q})^2}{8l^2( \ntwo{T}^2 + \ntwo{P^2})}\right]  \le \frac{1}{l^2},
$$
where we have used $\exp(-\ln^3 l) < \frac{1}{2(l^ 2)}$, for $l>10$.
}

\paragraph{Bringing it all together: main algorithm and analysis}
\begin{algorithm}[t]
\caption{Distinguishing between two distributions}
\label{alg:main}
\alginp Blackbox providing samples from $P$, $Q$ and $T$. Estimate  $s=\numsamples{P}{Q}$.\footnote{While we assume that
the separation between distribution is given, it can be estimated, by adding extra poly-logarithmic factor: run the algorithm several times
and check the consistency of the result. If the result is not consistent, double  $s$  and iterate.}
\algout ``P'' if $T$ is $P$ and ``Q''  otherwise \\
\algdesc
 \begin{enumerate}
 \item Compute $L_2$ norm of $P$, $Q$ and  $T$ using Algorithm~\ref{alg:2-norm},  using accuracy parameter $l=30 s \logth s$.
 Repeat $\log s$ times.  Let  $\tilde{P}_i$, $\tilde{Q}_i$ and $\tilde{T}_i$ denote the estimated norms in $i$th iteration.
 \item If $\tilde{T}_i\ge \tilde{P}_i$ for all $i$ or $\tilde{T}_i \le \tilde{P}_i$  for all $i$ then report ``Q'' and quit.
 \item If $\tilde{T}_i \ge \tilde{Q}_i$ for all $i$ or $\tilde{T}_i \le \tilde{Q}_i$  for all $i$ then report ``P'' and quit.
  \label{stage-1-finish}
\item Else:
\begin{enumerate}
  \item Training Phase: sample $l=30\logth{s}s$ elements from distributions $P$ and $Q$. Let $C_P$ and $C_Q$ denote the configuration of elements that were selected.
  \item Testing Phase: use Algorithm~\ref{alg:sampling} to sample with repetition $\nsamples $ times, on  both $C_P$ and $C_Q$ using fresh sample each time.
 Let $c_P$ and $c_Q$ denote the  total size of the Algorithm~\ref{alg:sampling} output for $C_P$ and $C_Q$ respectively.
  \item If $c_P > c_Q$ report ``P'', otherwise report ``Q''.
  \end{enumerate}
 \end{enumerate}
\end{algorithm}

Let $W_P$ and $W_Q$ denote the total probability mass of sample selected by P and Q in distribution $T$. In other words $W_P = \sum_{i=1}^n t_{s_P(i)}$,  where $s_P(i)$ is the $i$-th sample from $P$.

 Now, consider $H_1$ (e.g. $T=P$). We have:
$
 \Ex{W_P} = \nsamples \sum_{i=1}^n p_i ^2
$
and $\Ex{W_Q} = \nsamples \sum_{i=1}^n p_i q_i$, where $i$-th term is expected contribution
of $i$-th element of the distribution into the total sum for a single selection.
Therefore we have
\[
\Ex{W_P - W_Q} = \nsamples \sum_{i=1}^n p_i (p_i - q_i)
\]
Similarly in hypothesis $H_2$ we have:
$$
\Ex{W_Q - W_P} = \nsamples\sum_{i=1}^n {q_i} (q_i - p_i)
$$
The rest of the analysis proceeds as follows, we first show that if the algorithm passes the first stage then with high probability
$|(\ntwo{P}^2 - \ntwo{Q}^2)| \le 4 \logth{\nsamples} \ntwo{P+Q}$, in which case the majority voting on collision counts gives the desired result.
The following lemma is almost immediate from Lemma~\ref{lemma:algorithm-2-norm}
\begin{lemma}[Correctness of the case when $\ntwo{P}\not \approx\ntwo{Q}$]
\label{lem:correctness-approx}
   If $$\nsamples|\sum_{i=1}^n (p_i^2 - q_i^2)|
                  \ge \ln^{3/2}(l)( \ntwo{P}+\ntwo{Q})$$
the algorithm terminates
on or before step \ref{stage-1-finish} of algorithm \ref{alg:main} with probability at least $1-c/\nsamples^2$. Further,
the probability of incorrect answer if it terminates is at most $\frac{c}{\nsamples^2}$, for some constant $c$.
\end{lemma}
\proof{
Without loss of generality we can assume $T=P$. Using the result of lemma~\ref{lemma:algorithm-2-norm},  the probability
that $\tilde{P}_i<\tilde{T}_i$ or for all $i$ is at most $(\frac{1}{2})^{2\log \nsamples} \le \frac{1}{\nsamples^2}$. Thus
probability of incorrect answer is at most $\frac{1}{\nsamples^2}$.

Second, if the condition of the lemma satisfied, then from lemma~\ref{lemma:algorithm-2-norm} and the union bound the probability of inconsistent measurements is at most $\frac{\log l}{l^2} \le \frac{1}{\nsamples^2}$.
}

Given the lemma above, we have that if the algorithm passed stage 1, then
$$
  -4\logth{l}(\ntwo {P} +\ntwo{Q}) \le
    l(\sum p_i^2 - \sum q_i^2 )\le 4\logth{l}( \ntwo {P}+\ntwo{Q})
$$
Therefore in hypothesis $H_1$:
\[
\Ex{W_P - W_Q}  = l(\sum_{i=1}^n p_i^2   - \sum_{i=1}^n q_i p_i ) \ge
{l} \times \left[\frac{\ntwo{P-Q}^2}{2} - \frac{4\logth{l}}{l} (\ntwo {P}+\ntwo{Q})\right]
\]
\omt{
Similarly in $H_2$ we have:
$$
\Ex{W_Q - W_P} = {r} \times \left[\frac{\ntwo{P-Q}^2}{2} - \frac{4\logth{l}}{l} (\ntwo {P}+\ntwo{Q})\right]
$$
}

We have $\ln l = \ln (\nsamples \ln^{3/2} \nsamples) \le \ln \ln s + \log s + 3  \le 1.5 \log s$, and thus
$l \ge 20\logth{l}  \numsamples{P}{Q}$. Substituting we have:
$$
\Ex{W_P- W_Q} \ge l \left[\frac{\ntwo{P-Q}^2}{2}  - \frac{4}{20} \ntwo {P-Q}^2\right] \ge 6 \logth{l}(\ntwo{P} + \ntwo{Q})
$$
Applying Lemma~\ref{cor:weight-concentration} with weight function  $P$,
the probability that either $W_P$ and $W_Q$ deviates from its expectation by more than $2 \logth{l} \ntwo{P}$
is at most $\frac{c}{\nsamples^2}$  for a fixed constant $c$. Thus
\begin{equation}
\label{eq:estimates-are-close}
\Pr{W_P - W_Q  \le \logth{l}( \ntwo {P} + \ntwo{Q})} \le \frac{c_2}{\nsamples^2}
\end{equation}

Therefore $W_P$ and $W_Q$ with high probability are far apart,
and could be distinguished using the bounds from Lemma \ref{cor:bernoulli}.
Indeed, the  total expected number of hits is $sW_P$ and $sW_Q$,
for both $P$ and $Q$, thus the probability that the total
number of hits is for $C_P$ is smaller than $C_Q$ in $H_1$ is at most:
\begin{equation}
\label{eq:collisions-are-fine}
\Pr{c_p < c_q} \le \exp[-\frac {s  (W_P-W_Q)^2}{W_P + W_Q}]   \le
\exp\left [-\frac{s \ln^{3}l\ntwo{P+Q}^2 }{l(\ntwo{P}^2 + \ntwo{Q}^2)} \right] \le \frac{1}{l ^2}
\end{equation}
Combining~\eqref{eq:estimates-are-close} and~\eqref{eq:collisions-are-fine}
we have that  for some universal constant $c$, with probability at least $1-\frac{c}{\nsamples^2}$,   $C_P$ will receive more hits than $C_Q$ and symmetrically in $H_2$, $C_Q$ will receive more hits than $C_P$. Thus the algorithm produces correct answer with high probability and the proof of Theorem~\ref{thm:alg} is immediate.

\section{Lower Bounds for  Weakly Disjoint Distributions}
\label{sec:lower-bounds}
In this Section we prove Theorem~\ref{thm:lower-bounds}.
First we observe that for any {\em fixed} pair of distributions, there is in fact an algorithm that can differentiate between them with far fewer samples than our lower-bound theorem dictates\footnote{For instance distinguishing between two uniform distribution on half domain a constant number of samples is sufficient}. Thus,  the main challenge is to prove that there is no universal algorithm that can differentiate between arbitrary pairs of distributions.  Here, we show that even the simpler problem where the pair of distributions is fixed, and a random permutation $\pi$ is applied to both of them, there is no algorithm that can differentiate between $\pi P$, and $\pi Q$. Since this problem is simpler than the original problem (we know the distribution shape), the lower bound applies to the original problem.

\begin{problem}[Differentiating between two known distribution with unknown permutation]
\label{prb:unknown-perm} Suppose $P$ and $Q$ are two known distributions on domain $D$.
Solve distinguishability problem on the class of distributions defined by $(\pi P, \pi Q)$, for all permutations $\pi$.
\end{problem}
In Problem~\ref{prb:unknown-perm}, the algorithm needs to solve the problem for every $\pi$, thus if such algorithm exists, it would be able
to solve distinguishability problem with $\pi$ chosen at random and from the perspective of the algorithm, elements that were chosen the
same number of times, are equivalent. Thus, the only factor the algorithm can differentiate upon are counts of how often
each element appeared in different phases. More specifically we will use $\numsigs ijk$ notation to denote the number of elements, which
were chosen $i$-times while sampling from $P$, $j$-times while sampling from $Q$ in the training phase and $k$-times during the testing phase.
We also use notation $\numsigs ij*$ to denote the total number of elements that were selected $i$ and $j$ times during training phase and arbitrary number of times during the testing phase. Finally  and $\numsigs ij+$ to denote the number of elements that were selected at least once during the testing phase.
\omt{
Our main tool  for  the analysis is Neyman-Pearson lemma~\cite{neyman1933problem}, that allows us to construct the optimal algorithm and explore its properties.
}
In what follows we use $H_1$ and $H_2$ to denote the two possible hypotheses that the testing distribution is in fact $P$ or $Q$ respectively.

To prove this theorem we simplify the problem further  by disclosing some additional information about distributions,
this allows us to show that some data in the input possesses no information and could be ignored. Specifically, we rely on the
following variant of the problem:
\begin{problem}[Differentiating with a hint] Suppose $P$ and $Q$ are two known distributions on $D$ \label{prb:diff-hint}
and $\pi$ is unknown permutation on $D$.
For each element that satisfies one of the following conditions the algorithm is revealed whether it belongs to common or disjoint set.
(a) selected at least once while sampling from $T$ and at least twice while sampling from $P$ or $Q$
(b) selected at least twice while sampling from $P$ or $Q$, and belongs to the $ \common{\pi P, \pi Q}$/
The set of all elements for which their probabilities are known is called {\em hint}.
Given the hint, differentiate between $\pi P$ and $\pi Q$.
\end{problem}
Note that Problem \ref{prb:diff-hint} is immediately reducible to Problem \ref{prb:unknown-perm}, thus a lower bound for \ref{prb:diff-hint} immediately implies a lower bound for \ref{prb:unknown-perm}.
If an element from the disjoint part of $P$ and $Q$ has its identity revealed, then  the algorithm can immediately output the correct answer.  We call such elements {\it helpful}. We call other revealed elements {\it unhelpful}, note that the set that the unhelpful elements is fully determined by the training phase (these are the elements
that were selected at least twice during training phase and belong to the common set). First, we prove the bound on
the probability of observing helpful elements. Later we show that knowledge of unhelp does not reveal much to the algorithm.
\begin{lemma}
 \label{lemma:helpfuls} If the number of samples $s \le  \frac{0.25}{\nthree{P-Q}}$, then the probability that there is one or more helpful elements is
 at most $\onesixth$.
\end{lemma}
\proof{
For every element that has probability of selection  during testing phase $p$, the probability that it becomes helpful is at most $\binom{2s}{3}p^3 < \frac{8s^3p^3}{6} \le 1.5 s^3 p^3$
if it belongs to disjoint section, and 0 otherwise.  Therefore, the total expected number of helpful elements is $1.5s^3\nthree{P-Q}^3$. Using Markov inequality we immediately
have that probability of observing one or more helpful element is at most $\frac{1}{1.5s^3\nthree{P-Q}^3} \le \onesixth $ as desired.
}

Since the probability of observing helpful element is  bounded by
$\onesixth$, it suffices to prove that any algorithm that does not observe any {\em helpful} elements, is correct with probability at most $1- \Omega(1)$. The next step is to show that disclosed elements from the common part are ignored by the optimal algorithm.
\def\alg{{\cal A}}
\begin{lemma}
 \label{lemma:common-elements-nonhelpful}
   The optimal algorithm does not depend on the set of unhelpful elements.
\end{lemma}
\proof{
 More formally, let $\config$ denotes the testing configuration, and let $Y$ denotes
the unhelpful part of hint. Let $\alg(C, Y)$ is  the optimal algorithm that takes as that outputs $H_1$ or $H_2$. Suppose there exists
$Y'$ and $Y''$ such that $\alg(C, Y') \not = \alg(C, Y'')$, and without loss of generality we assume $\alg(C, Y')=H_1$ and $\alg(C, Y'')=H_2$.
Of all optimal algorithms we chose the one that minimizes the number of triples $(C, Y, Y')$ that satisfy this property. Without loss of generality
$\Pr{C | H_1} \ge \Pr{C | H_2}$ and let $\alg_1$ be the modification of $\alg$ such that $\alg_1(C, Y'') = H_1$.  But then the total probability of
error for $\alg_1$ will be the $\Pr{C | H_1}  Pr{Y''} - \Pr{C | H_2}\Pr{Y''}$, which is smaller or equal than $\alg$ contradiction with either
optimality  or minimility of $\alg$.
}

So far we showed that {\em helpful} elements terminate the algorithm, and {\em nonhelpful} do not affect the outcome. Therefore, the only signatures $(i, j, k)$ that
the algorithm has knowledge of will belong to
the following set: $$\{\sigs000, \sigs001, \sigs010, \sigs100,
\sigs011, \sigs101, \sigs002, \sigs200, \sigs020\}.$$
\omt{Further, the elements that have signatures
$\sigs 002, \sigs200, \sigs020$ all belong to the disjoint set $\disjoint{\pi P, \pi Q}$.}
Consider the following random variables $\numsigs01*$, $\numsigs10*$, $\numsigs00*$, $\numsigs02*$,$ \numsigs20*$.
They are fully determined by the training phase and thus
are independent of the hypotheses. We call these random variables the training configuration. The following lemma
is immediate and the proof is deferred to appendix.
\begin{lemma}
\label{lemma:ignore-sigs}
If the training configuration is fixed then the values $\numsigs 011$, $\numsigs101$, $\numsigs001$
fully determine all the data available to the algorithm.
\end{lemma}
Therefore for fixed configuration of the training phase, the output of optimal algorithm only
depends on $\numsigs 011$, $\numsigs101$, $\numsigs001$.
\newcommand{\hits}[2]{\ensuremath{  h(#1,#2)}}
Consider $\hits ij$ the total number of elements that were selected during testing phase, and that have signature \sigs ij*. Note that
$\hits ij = \sum_{k=1}^{\nsamples} k \times |\sigs ij k|$.

Now we prove that no algorithm by observing \hits 01, \hits 10,   \hits 00 can have error probability less than $1/17$. Again we
defer the proof to the appendix due to space constraints.
\begin{lemma}
\label{lem:lower-h}
   Let $C_P, D_P$ and $C_Q, D_Q$ be the probability masses of subsets of
$\common{\pi P, \pi Q}$ and $\disjoint{\pi P, \pi Q}$ that were
sampled during training phase of $P$ and $Q$ respectively. Assume without loss of generality
$\Pr{D_P \ge D_Q} \ge 1/2$.  Then with probability at least $1/2$, in hypotheses $H_1$ for observed $\hits 10=a$, $ \hits 01=b$
$\hits 0 0 = c$, we have:
\begin{equation}
 \frac{\Pr{~ \hits 10=a, \hits 01= b, \hits 00 = c | H_1}}
         {\Pr{~\hits 10=a, \hits 01 = b, \hits 00 = c | H_2}}  \le 8
\end{equation}
\end{lemma}
From this lemma it immediately follows that any algorithm will be wrong with probability at least $1/16$.
Now we are ready to prove the main theorem of this section.
\prevproof{theorem}{thm:lower-bounds}{
By lemmas \ref{lemma:common-elements-nonhelpful} and \ref{lemma:ignore-sigs}, the optimal algorithm can ignore all the elements that has signatures other than
$\numsigs011$, $\numsigs101$ and $\numsigs001$. Now, suppose there exists an algorithm ${\cal A}(x, y, z)$ that only observes $\numsigs 011$, $\numsigs 101$, $\numsigs001$, and has error probability
less then $1/100$. Then the algorithm for $\numsigs01+$, $\numsigs 10+$ and $\numsigs00+$ that just substitutes the latter into former, will have error probability at most $1/100 + \onesixth < 1/17$. Indeed, let $x=\numsigs 01+$, $y=\numsigs 10+$ and $z=\numsigs00+ -  2(\nsamples - x -y)$ and execute
the algorithm ${\cal A}(x, y, z)$ will have error at most $1/100 + \onesixth$.
Thus, we contradicted Lemma~\ref{lem:lower-h}.
}

\section{Conclusions and Open Directions}
Perhaps the most interesting immediate extension to our work is to incorporate high-frequency elements in our analysis to eliminate the technical assumptions that we make in our theorems theorem. One possible approach is to combine techniques of Valiant and Micali, but it remains to be seen if the
hybrid approach will produce better results. Proving or disproving our conjecture that weakly disjoint distribution are indeed the hardest when it comes to
telling distribution apart would also be an important technical result.

The other directions is to extend the techniques of section \ref{sec:preliminaries}. For instance this techniques could be used to estimate various concentration
bounds on how many heterogeneous bins will receive exactly t balls and  it remains an interesting question on how far those techniques could be pushed.
 On the  more technical side  an interesting question is whether (under some reasonable assumption), the probability ratio between type I and type II configurations
is, in fact bounded by constant, rather than by $O(\nsamples)$. By using tighter analysis we can  in fact show that this ratio is in fact $O(\sqrt{\nsamples})$ though reducing it further remains an open question.

\bibliographystyle{plain}
\bibliography{bibliography}
\appendix

\section{Equivalence of identity and distinguishability problems}
\prevproof{Lemma}{lemma:identity-distinguish}{
\def\Alg{\cal A}
We first show how to simulate a correct algorithm $\Alg_I$ for the identity problem using an algorithm $\Alg_D$ for the distinguishability problem.
We run  $\Alg_D$ for $3\log{s}$ times on fresh input, where the sample for the testing phase is always taken
from the first distribution. If the answer is always the same say ``different'', otherwise say the ``same''. If the two distributions are the same,
then $\Alg_D$ gives random answer, therefore the probability for $\Alg_D$ produce the same answer for $\log s$ iterations
 (and hence $\Alg_I$ producing wrong answer)  is
at most  $(3/5)^{3\log {s}} \le \frac{1}{s^2}$. Similarly if distributions are different, the probability for $\Alg_D$ go give different answer
 means that it has mistaken at least once, which would happen with probability at most $\frac{\log(3 s) \polylog (s)}{s^2}$ as desired as it remains polylograthmic.

For the other direction we simulate a correct algorithm for the
distinguishability problem  using an algorithm $\Alg_I$ for the identity problem. We
test a new sample against both $X$ and $Y$, if $\Alg_I$  says that both are the same or both are different,
output $X$ with probability 0.5.  If the output of $\Alg_I$ is ``the same as $X$'' output ``$X$'', otherwise output ``$Y$'''.

If $X$ and $Y$ are the same the testing algorithm will say ``the same'' with probability at least $1-2\polylog(s)/s^2$, thus distinguishing algorithm
will say $X$ with probability at least $0.5 - \polylog(s)/s^2$, and $Y$ with probability at least $0.5 - \polylog(s)/s^2$.
 If $X$ and $Y$ are different, then the testing algorithm will make a mistake with probability at most $2\polylog(s)/s^2$, and thus the identity algorithm will
 produce different answers, and the distinguishing algorithm will be correct with probability at least $1-2\polylog(s)/s^2$.
}

\section{Proofs From Subsection~\ref{sec:bernoulli}}
\begin{lemma}
\label{thm:bernoulli}
  Let $\{x_i\}$ and $\{y_i\}$ be a sequence of $N$ independent, though not necessarily identical Bernoulli random variables such that $\Ex{\sum x_i} = pN$ and
  $\Ex{\sum y_i} = qN$ and $pN < qN$. If  $N \ge \frac{8|\log 2\delta|(p+q)}{(p-q)^2},$ then
$$\Pr{\sum x_i \ge \sum y_i} \le \delta.$$
\end{lemma}
\proof{
Using Chernoff inequality for any $\alpha$ we have
\begin{equation}
\label{eq:cher1}
\Pr{\sum_{i=1}^n x_i \ge (1+\alpha)pN} < \frac{\exp[\alpha \times pN]  }{(1+\alpha)^{(1+\alpha)pN}}.
\end{equation}
On the other hand each for $y_i$ we have:
\begin{equation}
\label{eq:cher2}
\Pr{\sum_{i=1}^n y_i \le (1-\beta)qN} < \exp[-\frac{Nq \beta^2}{2} ].
\end{equation}
Choose $1-\beta= \frac{p+q}{2q}$, and $1+\alpha = \frac{p + q}{2p}$. Since $p<q$, $\beta,\alpha > 0$.

First consider the case where $p \le q / 8$. Thus substituting it into \eqref{eq:cher1} we get:
\begin{align*}
\Pr{\sum_{i=1}^n x_i \ge \frac{p+q}{2}N} & \le \frac{\exp[\frac{q-p}{2}N]}{(\frac{p+q}{2p})^{N\frac{p+q}{2}}} \le
\exp\left[N\frac{q-p}{2} -  \frac{3(p+q)}4 N\right]
\le \exp[-N(\frac{q}4 + \frac{5p}{4})] \\
&\le \exp[-N \frac{(q-p)}{4} ] \le \exp \left[-N \frac{(q-p)^2}{4(p+q)}\right]
\end{align*}
where we have used $(p+q)/(2p) \ge 4.5\ge e^{1.5}$. Similarly for \eqref{eq:cher2} we have:
\begin{equation}
\Pr{\sum_{i=1}^n y_i \le (1-\beta)qN} \le \exp\left[-N(p-q)^2/8q\right] \le \left[-N \frac{(q-p)^2}{8(p+q)}\right]
\end{equation}
If, on the other hand $q \ge p \ge q/8$, then $\alpha = (p+q)/2p - 1 \le 4$, and we can use the following variant of Chernoff bound:
\begin{equation}
\Pr{\sum_{i=1}^n x_i \ge N\frac{p+q}{2}} \le \exp[-pN\alpha^2/4] \le \exp[-N(q-p)^2/(4p)] \le \exp[-N(q-p)^2/2(p+q)]
\end{equation}
and similarly:
$$
\Pr{\sum_{i=1}^n y_i \le N\frac{p+q}{2}} \le \exp[-pN\beta^2] \le \exp[-N(q-p)^2/(2p)] \le \exp[-N(q-p)^2/(p+q)]
$$
Combining we have the desired result.
}
Lemma \ref{cor:bernoulli} immediately follows from lemma \ref{thm:bernoulli}

\section{Proofs for concentration bounds for balls and bins}
\prevproof{Lemma}{lemma:weight-concentration-base}{
Note that type II sampling is over $\nsamples$ elements whereas type I is over $\lprime$ elements.
We have, $$
\PRI{\config} = p_1^{i_1}p_2^{i_2}\dots p_n^{i_n} \frac{\lprime!}{i_1! i_2! \dots i_n!},$$
whereas
$$
 \PRII{\config} = \prod \frac{\nsamples!}{i_j! (\nsamples-i_j)!}p_j^{i_j} (1-p_j)^{\nsamples-i_j} = \frac{p_1^{i_1}p_2^{i_2}\dots p_n^{i_n}}{i_1!\dots i_n!}
  \prod \frac{\nsamples!}{(\nsamples-i_j)!} (1-p_j)^{(\nsamples-i_j)}$$

Recalling that $i_j \le \ln \nsamples$, we have $\nsamples^{i_j}\ge \frac{\nsamples!}{ (\nsamples-i_j)!} \ge (\nsamples-\ln \nsamples)^{i_j}$, and $\sum i_j = \lprime$ therefore we have:
\begin{equation}
{\lprime^\lprime}e^{\nsamples-\lprime} \ge \nsamples^{\lprime} \ge \prod \frac{\nsamples!}{(\nsamples-i_j)}
\ge (\nsamples-\ln \nsamples)^{\lprime}=\nsamples^\lprime(1-\frac{\ln \nsamples}{\nsamples})^{\lprime}\ge \lprime^{\lprime}e^{\nsamples-\lprime} \frac{1}{2 \nsamples}.
\end{equation}
where we have used $\nsamples^\lprime=\lprime^{\lprime}(1+\frac{\nsamples-\lprime}{\lprime})^\lprime$ and
$e^{\nsamples-\lprime}\ge (1+\frac{\nsamples-\lprime}{\lprime})^\lprime \ge \frac{e^{\nsamples-\lprime}}{2}$.
Substituting we have:
$$
\frac{\PRII{\config}}{\PRI{\config}} \ge \frac{\lprime^{\lprime}e^{\nsamples-\lprime} }{(2e\nsamples)\lprime!}\prod \frac{(1-p_j)^\nsamples}{(1-p_j)^{i_j}} \ge
 \frac{e^{\nsamples}}{9e\nsamples\sqrt{\nsamples'}}\exp\left[-\sum_{j=1}^n p_j (\nsamples+1)\right] \ge \frac{1}{10e^2\nsamples^{3/2}}
$$
where in the first transition we used the fact that $(1-x/(\nsamples+1))^\nsamples \ge e^{-x}$ for $x<1$ and $\nsamples>1$, and
the fact that $p_j (\nsamples+1)<1$ and finally Stirling formula $\lprime! \le 3 \sqrt{\lprime}{\lprime/e}^\lprime$.
As desired.
To get  the lower bound we have:
\begin{align}
\frac{\PRII{\config}}{\PRI{\config}} & \le \frac{\lprime^\lprime e^{\nsamples-\lprime}}{\lprime!}\prod (1-p_j)^{\nsamples-i_j}
    \le \frac{e^\nsamples}{2\sqrt{\lprime}} \prod (1-1/\nsamples)^{-i_j}(1-p_j)^\nsamples \\
&\le \frac{3e^\nsamples}{2\sqrt{\nsamples}} \exp\left[-\sum_{j=1}^n p_j \nsamples \right] \le \frac{3}{2\sqrt{\nsamples}}
\end{align}
where in the second transition we used $n! \ge 2 \sqrt{n}(n/e)^n$ and $p_j < 1/l$, in the third we have used $(1-1/\nsamples)^{-\nsamples} < 3$,
and $(1-x/\nsamples)^\nsamples \le e^{-x}$.
}
\prevproof{Lemma}{lem:outlier-prob}{Indeed, let $d_i$ denote the total number of times that sample $s_i$ was repeated in $\sample$.
Recall that we have $p_i \nsamples < 1/2$ for all $i$, and thus we can just uniformly upper
bound it:
\begin{align}
  \Pr{d_i \ge \ln \nsamples} & \le \max_{k} \Pr{c_k \ge \ln \nsamples} \le
  \max_k \frac{\exp\left[\ln \nsamples - p_i \nsamples \right] }
{\left[\frac{\ln \nsamples}{p_k \nsamples}\right]^{\ln \nsamples}}  \\
& \le \max_k \frac{\nsamples (p_k \nsamples)^{\ln \nsamples} } {\nsamples^{\ln \ln \nsamples }}
\le \left(\frac{1}\nsamples\right)^{\ln \ln \nsamples+1}.
\end{align}
using union bound over all $d_i$ gives us the desired result.
}
\prevproof{Lemma}{concentration-lemma}{
Let us denote the set of all configurations satisfying
 $|\sum_{i=1}^n \alpha_i c_i - \Ex{\sum_{i=1}^n \alpha_i c_i }| > r$, by
 $\cal D$.
For any  configuration $C \in {\cal D}$ such that $\max c_i \le \ln \nsamples$,
we can apply lemma \ref{lemma:weight-concentration-base}. And thus
$\PRI{C} \le 30 s^{3/2}\PRII{C}$. In addition the total probability mass of
all configurations $C$
such that $\max c_i > \ln \nsamples$ is at most $\frac{1}{s^{\ln\ln\nsamples}}$.
Therefore $\PRI{\cal D} \le 30 s^{3/2} \Pr{|W' - \Ex{W'}| > r} + \frac{1}{s^{\ln\ln\nsamples}}$ as desired.
}
\prevproof{Lemma}{cor:weight-concentration}{
Consider type II sampling with $s$ samples and let $W'$ be the total selection weight. Note that
$\Ex W = \Ex W' = s \ntwo{P}^2$.  Further, we have  $W'   = \sum_{i=1}^{\nsamples} c'_i p_i$, where $c'_i$ is the count of how many times
$i$-th element was chosen. The individual terms in the sum are independent, however they are not bounded, so we cannot use Hoeffding inequality\cite{hoeffding}.
Instead we consider $V'= \sum_{i=1}^{\nsamples} \min(c'_i, \ln s) p_i$.
Using Hoeffding inequality we have:
$$
\PRII{|V' - \Ex{V'}| \ge 1.5(\ln \nsamples)^{3/2} \ntwo{A} }\le \exp\left[\frac{-4.5 \ln^3 \nsamples \ntwo{A}^2}{(\ln s)^2\sum_{i=1}^{n} \alpha_i^2} \right]
\le \frac{1}{\nsamples^4}
$$
furthermore, because of the lemma \ref{lem:outlier-prob} we have:
$$\Ex{V' - W'} \le \Pr{V' \neq = W'}\ninf{A}\nsamples \le \frac{\ninf{A}\nsamples  }{\nsamples^{\ln\ln \nsamples} }\le \ntwo{A}
$$
and hence
$$
\Pr{|W' - \Ex W'| \ge  2(\ln \nsamples)^{3/2} \ntwo{A}}  \le \Pr{|W' - \Ex V'| \ge  1.5(\ln \nsamples)^{3/2} \ntwo{P}} \le \frac{2}{\nsamples^4}
$$
where we have used that $V' \not = W' \le \frac{1}{\nsamples^4})$.
Thus, using the concentration lemma \ref{concentration-lemma}, we have
$$
\Pr{|W - \Ex{W}| \ge 2(\ln \nsamples)^{3/2} \ntwo{P}} \le
   \Pr{|W' - \Ex W'|  2(\ln \nsamples)^{3/2} \ntwo{P}}  \times s^{3/2} +\frac{1}{\nsamples^{\ln \ln \nsamples}  } \le 1/s^2
$$
}

\section{Proofs for lower bounds}

\prevproof{Lemma}{lemma:ignore-sigs}{Indeed:
\begin{align*}
\numsigs 002 &= (\nsamples  - \numsigs101 - \numsigs 011 - \numsigs 001)/2  \\
\numsigs 100 &= \numsigs 10* - \numsigs101,~ \numsigs 010 = \numsigs 01* - \numsigs011 \\
\numsigs 000 &= \numsigs 00* - \numsigs 002 - \numsigs 001, ~~\numsigs 020 = \numsigs 02*, ~~ \numsigs 200  = \numsigs 02*
\end{align*}
}
\prevproof{Lemma}{lem:lower-h}{
 Denote $\Pr{\hits 10=a, \hits 01= b, \hits 00 = c | H_i}$ as $\pi_i$. We need to bound $\pi_1/\pi_2$.
  We have  $\pi_1 = (C_P + D_P)^a C_Q^b (1- C_P - C_Q - D_P)^c$   and
  $\pi_2 = (C_P)^a (C_Q+D_Q)^b (1- C_P - C_Q - D_Q)^c$
  therefore
 \begin{align}
  \frac{\pi_1}{\pi_2} & =  \left(1 + \frac{D_P}{C_  P}\right)^a \left(1 + \frac{D_Q}{C_Q}\right)^{-b} \left(1 + \frac{D_P - D_Q}{1-C_P-C_Q-D_P}\right)^{-c} \\
     & \le 2 \exp[\frac{D_P}{C_P}a - \frac{D_Q}{C_Q}b  -  \frac{D_P - D_Q}{1-C_P-C_Q-D_P} {c	}] \\
\end{align}
  Now, we replace $a = a_0 + \delta_a$ with $b=b_0 + \delta_b$ and $c=c_0 + \delta_c$,
 where $a_0 = (C_P + D_P) \nsamples$, $b_0 = C_Q\nsamples$,  and $c_0 = (1 - C_P - C_Q - D_P) \nsamples$ and we get:
 $$
\frac{\pi_1}{\pi_2} \le 2 \exp[\frac{D_P^2}{C_P}\nsamples]\exp[\frac{D_P}{C_P}\delta_a]
 \exp[\frac{D_Q}{C_Q}\delta_b]\exp[\frac{|D_P - D_Q|}{1-C_P-C_Q-D_P} {\delta_c}]
 $$
 Note $\Ex{\delta_a} = \Ex{\delta_b} = \Ex{\delta_c} = 0$, and with probability at least $0.5$,
 $\delta_a \le 3\sqrt{(C_P + D_P)\nsamples}$, $\delta_b \le 3\sqrt{C_Q\nsamples}$ and
 $\delta_c \le 3 \sqrt{(1-C_P-C_Q-D_P)\nsamples}$.
Substituting we have with high probability:
\begin{equation}
\frac{\pi_1}{\pi_2} \le 2 \exp[\frac{D_P^2}{C_P}\nsamples]
        \exp[\frac{3D_P\sqrt{\nsamples}}{\sqrt {C_P}}]
        \exp[\frac{3D_Q\sqrt{\nsamples}}{\sqrt {C_Q}}]
        \exp[\frac{3|D_P-D_Q|\sqrt{\nsamples}}{\sqrt{(1-C_P-C_Q-D_P)}} ]
        \label{eq:last-equation}
\end{equation}
Let $\ntwo{P_D}$ denotes the 2 norm of the disjoint part of $P$ and $Q$ , and $\ntwo{P_C}$ the the 2-norm of  common part $P$ (or $Q$).
We have $\ntwo{P-Q}^2 = \ntwo{P_D}^2 + \ntwo{P_Q}^2$ and $\ntwo{P+Q} \le \ntwo{P_C + P_D}$.

With high probability we have $D_P \le 2\ntwo{P_D}^2\nsamples \le \frac{\ntwo{P_C}}4$,
where we used concentration bounds and substituted
$\nsamples \le \frac{1}{10}\numsamples{P}{Q} \le \frac{\ntwo{P_C}} {8(\ntwo{P_D} )^2}$.
Similarly $D_Q \le \ntwo{P_C} / 4$ and  $C_P \ge \frac{\ntwo{P_C}^2\nsamples }{2}$, therefore we have
$$
  \frac{D_P^2\nsamples}{C_P} \le \frac{2\ntwo{P_C}^2\nsamples}{16\ntwo{P_C}^2\nsamples}  \le 1/10
$$
Substituting in \eqref{eq:last-equation} and using the fact the last exponent is $o(1)$, we get the desired result.
}
\end{document}